\documentclass[prstab,twocolumn,showpacs,preprintnumbers,amsmath,amssymb]{revtex4}
\usepackage{graphicx}
\usepackage{dcolumn}
\usepackage{bm}
\usepackage[dvips]{epsfig}
\usepackage{subfigure}
\newcommand{\half}{\mbox{${\textstyle \frac{1}{2}}$}}
\newcommand{\bmath}[1]{\mbox{\boldmath ${#1}$}}

\def\fmn#1#2{\mbox{${\textstyle \frac{#1}{#2}}$}}
\newcommand{\ww}{\mbox{$\,$}}
\begin{document}
\title{Determination of Deuteron Beam Polarizations at COSY}
\author{D.~Chiladze$^{1,2}$}\email{d.chiladze@fz-juelich.de}
\author{A.~Kacharava$^{3,2}$}
\author{F.~Rathmann$^{1}$}
\author{C.~Wilkin$^{4}$}
\author{S.~Barsov$^5$}
\author{J.~Carbonell$^6$}
\author{S.~Dymov$^7$}
\author{R.~Engels$^1$}
\author{P.D.~Eversheim$^{8}$}
\author{O.~Felden$^{1}$}
\author{R.~Gebel$^{1}$}
\author{V.~Glagolev$^{9}$}
\author{K.~Grigoriev$^{1,5}$}
\author{D.~Gusev$^{7}$}
\author{M.~Hartmann$^{1}$}
\author{F.~Hinterberger$^{8}$}
\author{V.~Hejny$^{1}$}
\author{A.~Khoukaz$^{10}$}
\author{I.~Keshelashvili$^{1,2}$}
\author{H.R.~Koch$^{1}$}
\author{V.~Komarov$^{7}$}
\author{P.~Kulessa$^{1,11}$}
\author{A.~Kulikov$^{7}$}
\author{A.~Lehrach$^{1}$}
\author{B.~Lorentz$^{1}$}
\author{G.~Macharashvili$^{7,2}$}
\author{R.~Maier$^{1}$}
\author{Y.~Maeda$^{12}$}
\author{R.~Menke$^{10}$}
\author{T.~Mersmann$^{10}$}
\author{S.~Merzliakov$^{1,7}$}
\author{M.~Mikirtytchiants$^{1,5}$}
\author{S.~Mikirtytchiants$^{5}$}
\author{A.~Mussgiller$^{1}$}
\author{M.~Nioradze$^{2}$}
\author{H.~Ohm$^{1}$}
\author{D.~Prasuhn$^{1}$}
\author{H.~Rohdje{\ss}$^{8}$}
\author{R.~Schleichert$^{1}$}
\author{H.~Seyfarth$^{1}$}
\author{E.~Steffens$^{3}$}
\author{H.J.~Stein$^{1}$}
\author{H.~Str\"oher$^{1}$}
\author{S.~Trusov$^{13,7}$}
\author{K.~Ulbrich$^{8}$}
\author{Yu.~Uzikov$^{7}$}
\author{A.~Wro\'{n}ska$^{1,11}$}
\author{S.~Yaschenko$^{3,7}$}

\affiliation{$^{1}$Institut f\"ur Kernphysik, Forschungzentrum J\"ulich, 52428 J\"ulich, Germany}
\affiliation{$^{2}$High Energy Physics Institute, Tbilisi State University, 0186 Tbilisi, Georgia}
\affiliation{$^{3}$Physikalisches Institut II, Universit\"at Erlangen--N\"urnberg, 91058 Erlangen, Germany}
\affiliation{$^{4}$Physics and Astronomy Department, UCL, London WC1E 6BT, U.K.}
\affiliation{$^{5}$High Energy Physics Department, Petersburg Nuclear Physics Institute, 188350 Gatchina, Russia}
\affiliation{$^{6}$Laboratoire de Physique Subatomique et de Cosmologie, 38026 Grenoble Cedex, France}
\affiliation{$^{7}$Laboratory of Nuclear Problems, Joint Institute for Nuclear Research, 141980 Dubna, Russia}
\affiliation{$^{8}$Helmholtz Institut f\"ur Strahlen und Kernphysik, Universit\"at Bonn, 53115 Bonn, Germany}
\affiliation{$^{9}$Laboratory of High Energies, Joint Institute for Nuclear Research, 141980 Dubna, Russia}
\affiliation{$^{10}$Institut f\"ur Kernphysik, Universit\"at M\"unster, 48149 M\"unster, Germany}
\affiliation{$^{11}$Institute of Nuclear Physics, 31342 Cracow, Poland}
\affiliation{$^{12}$Institut f\"ur Kernphysik, Universit\"at zu K\"oln, 50937 K\"oln, Germany}
\affiliation{$^{13}$Institut f\"ur Kern- und Hadronenphysik, Forschungszentrum Rossendorf, 01314 Dresden, Germany}

\date{\today}
\begin{abstract}
The vector ($P_z$) and tensor ($P_{zz}$) polarizations of a
deuteron beam have been measured using elastic deuteron--carbon
scattering at 75.6{\ww}MeV and deuteron--proton scattering at
270{\ww}MeV. After acceleration to 1170{\ww}MeV inside the COSY
storage ring, the polarizations of the deuterons were remeasured
by studying the analyzing powers of a variety of nuclear
reactions. For this purpose a hydrogen cluster target was employed
at the ANKE magnetic spectrometer, which is situated at an
internal target position in the ring. The overall precisions
obtained were about 4\% for both $P_z$ and $P_{zz}$. Though all
the measurements were consistent with the absence of
depolarization during acceleration, only an upper limit of about
6\% could be placed on such an effect.
\end{abstract}
\maketitle

\section{Introduction}
The COoler SYnchrotron (COSY) of the Forschungszentrum J\"ulich
(FZJ)~\cite{COSY} accelerates and stores protons up to
2.88{\ww}GeV and deuterons up to 2.23{\ww}GeV both for experiments
internal to the ring and for those using an extracted beam. In the
case of protons, an extensive series of successful measurements
with polarized beams and targets was carried out with the internal
EDDA detector~\cite{EDDA} and, used as a polarimeter, it can
determine the polarization of the beam with high precision. COSY
is now embarking on a program of
investigations~\cite{glagolev,STORI,SPIN} with polarized deuteron
beams and polarized deuterium storage cell targets~\cite{PIT} at
the ANKE magnetic spectrometer~\cite{ANKE}, placed at an internal
target station of the COSY ring~\cite{cosyring}. Much valuable
work has already been carried out to investigate the spin
manipulation of the internal deuteron beams in
COSY~\cite{morozov}. However, in order to accomplish the stated
program~\cite{SPIN}, it is a priority to establish polarization
standards for the deuteron beams, which are already available at
COSY, to better than 5\%. Higher precision than this has recently
been achieved in ANKE for polarized proton beams~\cite{yaschenko}.

Due to their much smaller anomalous magnetic moment, deuterons,
unlike protons, do not have to cross any first--order depolarizing
resonances while being accelerated in COSY, where the typical
working point tunes are $3.52$ through
$3.64$~\cite{Lehrach,morozov}. Even in the case of protons, higher
order resonances do not lead to any significant degradation. It is
therefore expected that there should be little or no loss of
polarization during acceleration and this has indeed been the
experience over many years at the SATURNE synchrotron, which
worked over a similar energy range~\cite{Arvieux}. Nevertheless,
it is important to check that this is true at COSY. This has been
done by measuring the beam polarization before and during the
acceleration, using calibrated polarimeters, and then deriving the
analyzing powers for four distinct nuclear reactions at
1170{\ww}MeV that were studied simultaneously with the ANKE
spectrometer: $\vec{d}p \to ^3$He$\,\pi^0$, $\vec{d}p$ elastic
scattering, $\vec{d}p \to (pp)n$, and quasi--free $\vec{n}p \to
d\pi^0$. In all these cases consistency was found with previously
published data. This confirms that we can use the Low Energy
Polarimeter (LEP) to monitor accurately the beam polarization
though, for its accurate determination at high energies, greater
reliance would be placed upon measurements within ANKE itself. The
secondary calibration standards established at 1170{\ww}MeV  can,
when necessary, be exported to other energies, as is routinely
done for protons~\cite{pollock,yaschenko}.

In Sec.~\ref{chap:ion-source} we describe the construction and
operation of the polarized ion source, which is based upon the
charge exchange with cesium atoms. In the configuration of the
radiofrequency transitions used here, the polarized ion source
provides one unpolarized plus seven other states. For each of
these states there is an optimal beam vector and tensor
polarization. However, these are never reached in practice and the
actual values have to be established using a beam polarimeter. The
deuteron beam is injected into the COSY ring at 75.6{\ww}MeV. The
LEP of Sec.~\ref{chap:beam-polarimeter}, which is based upon
scattering from a carbon target, is used to measure the beam
polarization at this energy. Unfortunately, the LEP is only
sensitive to the vector polarization of the beam, but this
drawback is well made up by measurements with the EDDA
polarimeter~\cite{EDDA} using elastic deuteron--proton scattering
at 270{\ww}MeV, for which accurate analyzing powers are
available~\cite{sekiguchi}. Consistency is found state--by--state
for these two lower energy measurements, though the calibration
standard is more precise for the 270{\ww}MeV data.

Only the principal elements of the ANKE spectrometer, including
the Silicon Tracking Telescopes (STT) used to measure very slow
particles emerging from the thin targets, are discussed in
Sec.~\ref{chap:ANKE}. The following section
\ref{Anke_measurements} describes the study of the four nuclear
reactions of interest in ANKE and shows that the results obtained
from the two measurements of the vector analyzing power and three
of the tensor are in complete agreement with previously published
results. Our conclusions are drawn in Sec.~\ref{chap:conclusions}.

\section{Polarized Ion Source\label{chap:ion-source}}
\subsection{Set--up of the Source}
\begin{figure}[htb]
\includegraphics[width=\columnwidth]{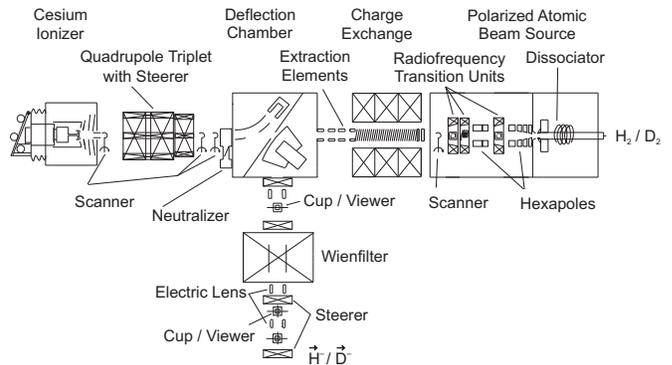}
\caption{Set--up of the polarized ion source at COSY.} \label{PIS}
\end{figure}

The polarized colliding beams source at
COSY~\cite{Gebel1,Gebel2,Gebel3} comprises three major groups of
components, the pulsed atomic beam source, the cesium beam source,
and the charge--exchange and extraction region. The set--up is
shown schematically in Fig.~\ref{PIS}.

The atomic beam source produces an intense pulsed polarized atomic
hydrogen or deuterium beam. The gas molecules are dissociated in
an inductively coupled rf discharge and a high degree of
dissociation is maintained by having a special admixture of small
amounts of nitrogen and oxygen that reduces surface and volume
recombination. The current output of the source depends
sensitively on the relative fluxes of the gases and on their
timing with respect to the dissociator radio frequency.  The atoms
are cooled to about 30{\ww}K by passing through an aluminum nozzle
of 20{\ww}mm length and 3{\ww}mm diameter and the resulting beam
is focused by an optimized set of permanent hexapoles into the
charge--exchange region. By cooling the supersonic atomic beam,
the acceptance of the hexapole system and the dwell time in the
charge--exchange region are increased, though scattering in the
vicinity of the nozzle reduces partially these beneficial effects.
A peak intensity of $7.5\times10^{16}${\ww}atoms/s has been
measured within a diameter of 10{\ww}mm at the exit of the
hexapole chamber.

The atomic $\vec{\rm H}^{\circ}$ beam with high nuclear
polarization collides with the fast neutral $\rm {Cs}$ beam inside
the charge--exchange region and an electron is transferred through
the $\vec{\rm H}^{\circ} + {\rm Cs}^{\circ} \to \vec{\rm H}^{-} +
{\rm Cs}^{+}$ reaction. The $\vec{\rm H}^{-}$ ions are extracted
from this region by electric fields before being deflected
magnetically through $90^{\circ}$, subsequently they pass a Wien
filter that provides the proper spin alignment for injection into
the cyclotron JULIC, which is the pre--accelerator for COSY.

The fast neutral cesium beam is produced in a two--step process.
Cesium vapor is thermally ionized on a hot porous tungsten surface
at a beam potential around 45{\ww}kV and the beam is then focused
by a quadrupole triplet to the charge--exchange region. The
space--charge compensation of the intense beam is improved by
feeding $10^{-3}${\ww}mbar$\ell$s$^{-1}$ argon to the beam tube
following the extraction. Cesium sputtering and contamination
generally impedes long--term reliability so that pulsed operation
of the cesium ionizer has been included in the
source~\cite{Gebel4}. The cesium pulses reach peak intensities of
over 10{\ww}mA with a width of about 10{\ww}ms. For routine
operation, cesium pulses with a 5{\ww}mA flat shape of 20{\ww}ms
width and a repetition rate of 0.5{\ww}Hz are used~\cite{Gebel3},
matched to the COSY injection scheme.  A neutralizer is placed
between the quadrupoles and the cesium deflector. This consists of
a cesium oven, a cell filled with cesium vapor, and a magnetically
driven flapper valve between the oven and the cell. The remaining
$\rm {Cs}^{+}$ beam is deflected in front of the solenoid to the
cesium cup. Routinely, a neutralizer efficiency of over 90\% is
achieved.

The highly selective charge--exchange ionization produces only
little unpolarized background that would reduce the nuclear beam
polarization. In the charge--exchange region, various beam
properties can be adjusted. Transverse emittance can be traded for
polarization by varying the solenoid's magnetic field. The field
strength during normal operation is 1.8{\ww}kG. The magnitude of
the electrical drift field inside the solenoid can be tuned to
optimize the energy spread of the beam. The electric field
gradient amounts to 0.5--1.0{\ww}V$\rm m^{-1}$. A monotonic gradient, 
in combination with a double buncher system in the injection beam
line to the cyclotron, leads to an improved bunching factor of
about four, compared to a factor of two for unpolarized beams.

Without modification of the system, the colliding--beams ion
source can provide negatively--charged polarized hydrogen and
deuterium beams of comparable intensities. To prepare polarized
deuterons with the desired combinations of vector and tensor
polarization, the atomic beam part of the source is equipped with
new radiofrequency transitions (RFTs). These transition units are
operated at the magnetic fields and radiofrequencies that allow
exchange of occupation numbers of the different hyperfine states
in deuterium~\cite{haeberli}. A set of three installed devices,
RFT$_1$ to RFT$_3$, allows a large number of combinations to be
delivered to experiments, as described in the following section.

\subsection{Operation of the Source}
The polarized $\rm H^{-}$ or $\rm D^{-}$ ion beam delivered by the
source, is pre--accelerated in the cyclotron JULIC and injected by
charge exchange into the COSY ring. The acceleration of vertically
polarized protons and deuterons at COSY is discussed in detail for
example in Ref.~\cite{Lehrach}.

The scheme used to produce the polarized deuteron beam consists of
eight different states, including one unpolarized state and seven
combinations of vector and tensor polarizations, obtained by
switching on or off the three radiofrequency transitions RFT$_i$,
where $i=1,\,2,\,3$. The states and the nominal (ideal) values of
the polarizations ($P_z$ and $P_{zz}$) and relative intensities
($I_0$) are listed in Table~\ref{tableLEPEDDA}.  The polarized ion
source was switched to a different polarization state for each
injection into COSY, in order to reduce the systematic errors. The
duration of a COSY cycle was sufficiently long ($\approx
200${\ww}s) to ensure stable conditions for the injection of the
next state \cite{rf-switch}. After the seventh state, the source
was reset to the zeroth mode and the pattern repeated for the next
injection. The ANKE data acquisition system received status bits
from the source, latched during injection. This ensured the
correct identification of the polarization states during the
experiment.

\section{Beam Polarimetry\label{chap:beam-polarimeter}}
\subsection{Low Energy Polarimeter}
To assist in the optimization of the polarization of the beams
inside COSY, a Low Energy Polarimeter (LEP, located in the
injection beam line) consisting of a UHV chamber with eight
flanges covered with thin stainless steel foils has been
used~\cite{cracow}. The moveable target frame is equipped with
viewers, allowing adjustment of the beam position. Several carbon
targets can be used for polarimetry measurements based on $d$C
elastic scattering. It is possible to place detectors at azimuthal
angles $\phi=0^{\circ}$, $90^{\circ}$, $180^{\circ}$ and
$270^{\circ}$ in the ranges of polar angles $25^{\circ}$ to
$70^{\circ}$ and $110^{\circ}$ to $155^{\circ}$.  NaI
scintillators, directly coupled to photomultipliers, are used for
particle identification. A set of exchangeable apertures, placed
in front of the scintillators, defines the acceptance of the
detector to be $\pm0.4^{\circ}$ for the measurements discussed
here.

The LEP is used at the COSY injection energy of $T_d=75.6${\ww}MeV
($p_d=539${\ww}MeV/c). Studies of the cross section, analyzing
power $A_y$, and the resulting figure of merit for
70{\ww}MeV~\cite{Kato} and 76{\ww}MeV~\cite{Stephenson} deuterons
suggest that the polarimeter should work best if the detectors are
placed to accept polar angles near $40 ^{\circ}$. Unfortunately,
under such conditions the tensor analyzing powers are very small
for this reaction so that the LEP is only sensitive to the vector
polarization of the beam~\cite{improvement}. Taking the two
data sets~\cite{Kato,Stephenson} together, we deduce that
$A_y(40^{\circ})=0.61\pm 0.04$ at 75.6{\ww}MeV.

\begin{table*}[htb]
\renewcommand{\arraystretch}{1.6}
\begin{small}
\begin{center}
\begin{tabular}{|c|c|c|c|c|c|c||c|c|c|c|}
\hline Mode & $P_z^{\mathrm{Ideal}}$ & $P_{zz}^{{\rm Ideal}}$ &
$I_0^{{\rm Ideal}}$ & RFT$_1$ & RFT$_2$ & RFT$_3$ & $P_z^{{\rm
LEP}}$ & $P_z^{\mathrm{LEP}}/P_z^{\mathrm{Ideal}}$ & $P_z^{{\rm
EDDA}}$ & $P_{zz}^{{\rm EDDA}}$\\
\hline
\hline
$0$  &\phantom{-}0 &\phantom{-}0  &1  &Off     &Off     &Off    &$\phantom{-}0.000
\pm 0.010$ &---&$0$                        &$0$\\
$1$  &--\fmn{2}{3}&\phantom{-}0&1   &Off&Off&On   &$-0.516 \pm 0.010 $            &$0.774\pm0.015$
&$-0.499\pm0.021$           &$\phantom{-}0.057\pm0.051$\\
$2$  &+\fmn{1}{3}&+1&1           &Off&On &Off  &$\phantom{-}0.257 \pm 0.010$   &$0.771\pm0.030$
&$\phantom{-}0.290\pm0.023$ &$\phantom{-}0.594\pm0.050$\\
$3$ &--\fmn{1}{3}&--1&1         &Off&On &On   &$-0.272 \pm 0.010 $            &$0.817\pm0.030$
&$-0.248\pm0.021$           &$-0.634\pm0.051$\\
$4$ &+\half&--\half&\fmn{2}{3}  &On &On &Off   &$\phantom{-}0.356 \pm 0.013 $  &$0.712\pm0.025$
&$\phantom{-}0.381\pm0.027$ &$-0.282\pm0.064$\\
$5$ &--1&+1&\fmn{2}{3}          &On &On &On   &$-0.683 \pm 0.013$             &$0.683\pm0.013$
&$-0.682\pm0.027$           &$\phantom{-}0.537\pm0.064$\\
$6$ &+1&+1&\fmn{2}{3}           &On &Off&Off   &$\phantom{-}0.659  \pm  0.013$ &$0.659\pm0.013$
&$\phantom{-}0.764\pm0.027$ &$\phantom{-}0.545\pm0.061$\\
$7$ &--\half&--\half&\fmn{2}{3} &On &Off&On   &$-0.376 \pm 0.013$             &$0.752\pm0.027$
&$-0.349\pm0.027$           &$-0.404\pm0.065$\\
\hline
\end{tabular}
\end{center}
\end{small}
\caption{The table lists the eight configurations of the polarized
deuteron ion source, showing the ideal values of the vector and
tensor polarizations and the relative beam intensities obtained by
operating the three radiofrequency transitions (RFTs). Also shown
are the measured vector and tensor polarizations of the deuteron
beam with statistical errors. The determinations of $P_z^{\rm
LEP}$ were carried out at a momentum of 539{\ww}MeV/c using the
Low Energy Polarimeter (LEP), the ratio of these to the ideal
values are also given. The EDDA values of $P_{z}^{{\rm EDDA}}$ and
$P_{zz}^{\rm EDDA}$ were obtained at 1042{\ww}MeV/c, assuming that
state--0 was unpolarized. The systematic uncertainties of the
polarizations $P_z^{{\rm EDDA}}$ and $P_{zz}^{{\rm EDDA}}$,
employed in the subsequent analysis, amount to $\pm 0.04$.
\label{tableLEPEDDA}}
\end{table*}

After pulse height analysis of the detector signals, the recorded
spectra were fitted with a Gaussian plus a small residual linear
background. The LEP set--up is operated with appropriate intensity to
ensure that the dead--time of the data acquisition system (DAQ) is
negligible. With the stable spin axis of the beam oriented along
the $y$--direction, the number of particles scattered through a
polar angle $\theta$ and an azimuthal angle $\phi$, after
corrections for beam luminosity, can be written as~\cite{Ohlsen}
\begin{widetext}
\begin{equation}
\label{eq1}
N(\theta,\phi)=N_0(\theta)\,\left[1+\fmn{3}{2}
P_{z}A_{y}(\theta)\cos\phi
+\fmn{1}{4}P_{zz}\left\{A_{yy}(\theta)(1+\cos2\phi)
+A_{xx}(\theta)(1-\cos2\phi)\right\}\right]\:.
\end{equation}
\end{widetext}
The beam vector and tensor polarizations, $P_z$ and $P_{zz}$, are
labeled conventionally in the reference frame of the source,
whereas the $\vec{d}\textrm{C}\to d\textrm{C}$ vector and tensor 
analyzing powers, $A_y$, $A_{yy}$, and $A_{xx}$, refer to the 
reaction frame, where $x$ and $y$ lie respectively within and 
perpendicular to the plane of the COSY ring.

Confining to the case of right ($R$) and left ($L$) counters
placed at $\phi=0^{\circ}$ and $180^{\circ}$, respectively, this
reduces to:
\begin{eqnarray}
\nonumber
N_L(\theta)& = &N_{0}(\theta)\left[1+\fmn{3}{2} P_{z}A_{y}(\theta)
+\fmn{1}{2}P_{zz}A_{yy}(\theta)\right]\\
N_R(\theta)& = &N_{0}(\theta)\left[1-\fmn{3}{2} P_{z}A_{y}(\theta)
+\fmn{1}{2}P_{zz}A_{yy}(\theta)\right]
\end{eqnarray}
Using the measured values of $A_{yy}$~\cite{Kato}, together with
an expected tensor polarization of the deuteron beam of
$P_{zz}\approx 0.6$, it is seen that the contribution of a tensor
polarized beam to the number of scattered particles is on the
percent level so that, to the desired level of accuracy, we can
take
\begin{equation}
P_z=\frac{2}{3}\frac{1}{A_y}\left(\frac{N_L-N_R}{N_L+N_R}\right)\:\cdot
\end{equation}

The results of the $P_z$ measurements with the LEP for the
different states are shown in Table~\ref{tableLEPEDDA}. Also given
are the ratios of $P_z$ to the ideal polarization that could be
provided by the source for that state. The variation from 66\% to
82\% depends, among other things, on the number of RFTs involved,
as indicated in the table.

\subsection{EDDA Polarimeter}
The EDDA detector has been used to provide a wealth of high
quality polarized proton--proton elastic scattering data over a
wide range of energies (0.5--2.5{\ww}GeV) by using a thin internal
target and measuring during the energy ramp of the COSY
accelerator~\cite{EDDA}. With the same apparatus, elastic
scattering of polarized deuterons from hydrogen was studied at
$T_d=270${\ww}MeV ($p_d=1042${\ww}MeV/c)~\cite{EPAC}, where
precise values are known for both tensor and vector analyzing
powers~\cite{sekiguchi}. In this way values of both vector and
tensor polarizations of the circulating deuteron beam could be
obtained at this energy.

\begin{figure}[htb]
\begin{center}
\includegraphics[width=\columnwidth,clip]{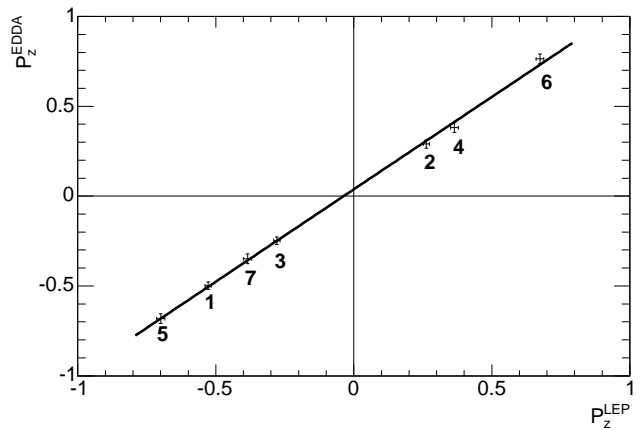}
    \caption{Comparison of the measurements of the vector polarization
    of the deuteron beam from EDDA and LEP for the seven states of the
    source, listed in Table~\ref{tableLEPEDDA}. The best fit straight
    line to the data is also shown.\label{LEPvsEDDA}}
\end{center}
\end{figure}

A fit to the data with the polarizations for all eight states
being left as free parameters yields $P_z=-0.002\pm0.038$ for
state--0. Any non--zero result might reflect a residual
polarization of state--0 or could be due to an instrumental
asymmetry, \emph{e.g.}\ caused by detector efficiencies. The data
cannot distinguish between these two possibilities. Therefore, the
EDDA values for the polarizations of the seven states shown in
Table I, were extracted under the assumption that state--0 is
unpolarized. Although supported by direct measurements with the
LEP, the uncertainty of about $\pm 0.04$ has to be considered as a
systematic uncertainty on all the polarizations extracted using
EDDA.

Since the EDDA and LEP data sets were taken with the same
conditions in the source, in order to determine the systematic
uncertainty of the polarizations, we compare quantitatively the
two sets of results for $P_z$. This is done for the seven states
in Fig.~\ref{LEPvsEDDA} using the data of
Table~\ref{tableLEPEDDA}. A linear fit of the two sets of results
with $\chi^2/\textrm{ndf}=5.1/5$ gives $P_z^{\rm EDDA} = (1.05\pm
0.02) P_z^{\rm LEP} +(0.038\pm 0.008) $. 
The value of the offset constant is compatible with the precision
of the EDDA calibration, as shown by the $\pm 0.04$ error bar in
the polarization of state--0. The total systematic error of the
slope in Fig.~\ref{LEPvsEDDA} amounts to $\pm 0.06$, where the accuracy of the
$dC$ analysing powers and the EDDA polarization determination has
been taken into account. Within these uncertainties, the observed
slope is consistent with unity. The typical fractions of the 
ideal vector and tensor polarizations were 74\% and 59\% respectively, 
though values from the individual polarization states were used 
in the subsequent analysis of the ANKE results.

\section{Experimental Set--up\label{chap:ANKE}}
\subsection{ANKE Magnetic Spectrometer}
\begin{figure*}[htb]
\begin{center}
\includegraphics[width=0.75\textwidth]{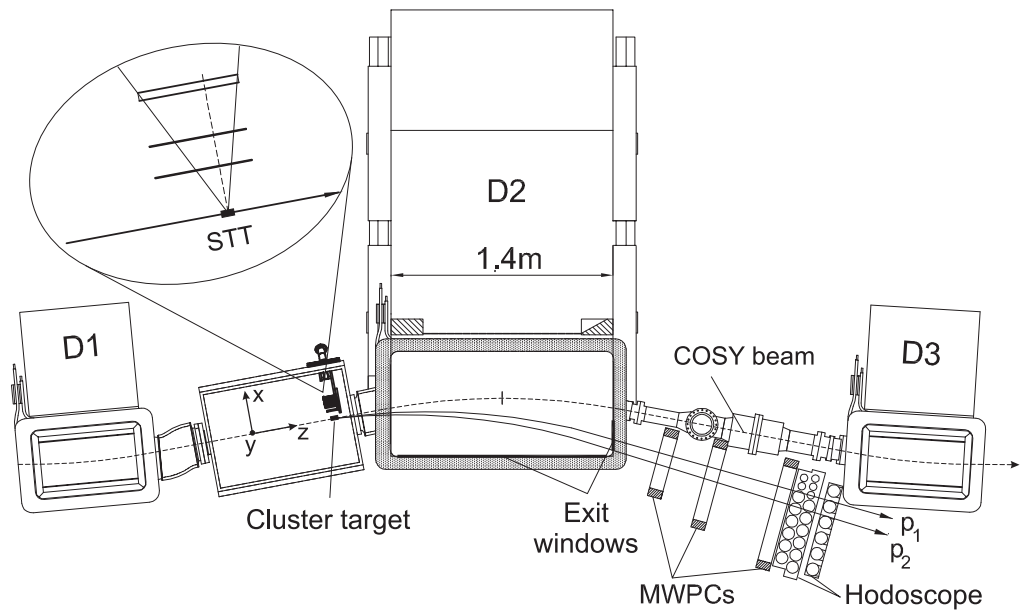}
    \caption{The ANKE experimental set--up, comprising the three
    dipole magnets D1, D2, and D3. Inside the target chamber, a
    Silicon Tracking Telescope (STT, shown in the inset and separately
    in Fig.~\ref{f:SPEC}) is mounted near the target jet, with the
    hydrogen clusters proceeding parallel to the $-y$ direction. The
    Forward Detector (FD) comprises three MWPCs and a hodoscope
    composed of three layers of scintillation counters.}
    \label{ANKEfig}
\end{center}
\end{figure*}

After being accelerated in the COSY ring~\cite{COSY}, the values
of the deuteron beam polarizations provided by EDDA and LEP can be
checked by measuring various nuclear reactions at
$T_d=1170${\ww}MeV ($p_d=2400${\ww}MeV/c) using the ANKE
spectrometer. This apparatus is described in Ref.~\cite{ANKE} and
we shall merely discuss the details of its additional features.
Thus Fig.~\ref{ANKEfig} shows only those parts of the spectrometer
that are relevant for the present experiment. The hydrogen
cluster--jet target of areal density $(3-5)\times
10^{14}${\ww}atoms/cm$^2$~\cite{khoukaz}, combined with an
internal beam of about $3\times 10^{9}$ stored vertically
polarized deuterons, provides a luminosity of up to
$10^{30}\,$cm$^{-2}$s$^{-1}$.

The reactions that are pertinent to this polarization study are:
\begin{itemize}
\item[] $\vec{d}p\to \,^3\textrm{He}\,\pi^0$,
\item[] Quasi--free $\vec{n}p\to d\pi^0$ with a fast
  spectator proton,
\item[] $\vec{d}p\to (pp)n$ producing a fast pair of
  protons with low excitation energy, and
\item[]$\vec{d}p\to dp$ at small angles.
\end{itemize}
The first three reactions can be measured using foremost
information from the ANKE Forward Detector (FD)
system~\cite{chil,dymov}. This comprises a set of three
multi--wire proportional chambers (MWPCs) and a three--plane
scintillation hodoscope, consisting of vertically oriented
counters (8 in the first plane, 9 in the second, and 6 in the
third). The third layer was implemented mainly to identify $^3$He.
The hodoscope system is capable of detecting also pairs of
particles, such as the protons emerging from the deuteron
charge--exchange reaction $dp \to (pp)n$ ~\cite{komarov}. Though
deuteron--proton elastic scattering can also be identified largely
by using the FD information, coincidence measurements with the
slow recoil proton being detected in a Silicon Tracking Telescope
(STT) yields more precise information.

\subsection{Silicon Tracking Telescope}
\begin{figure}
\begin{center}
\epsfig{file=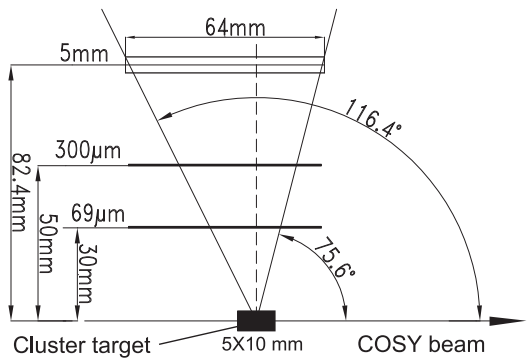,width=0.8\columnwidth} \caption{Top
view of the Silicon Tracking Telescope (STT), consisting of three
layers of different thickness. The approximate extension of the
cluster target beam in the $x$--$z$ plane is indicated, along with
the approximate polar angles covered.} \label{f:SPEC}
\end{center}
\end{figure}

For the identification and tracking of slow recoil protons, a
Silicon Tracking Telescope (STT) has been
developed~\cite{schleichert} that can be operated inside the
ultra--high vacuum of the accelerator.  The basic detection
concept of the STT combines proton identification with tracking
over a wide range in energy.  The tracking is accomplished by
three layers of double--sided micro--structured silicon strip
detectors that can be placed close to the target inside the vacuum
chamber (see Fig.~\ref{ANKEfig}).  The set--up of the STT is shown
in Fig.~\ref{f:SPEC}.
Measuring the energy loss in the individual layers allows
identification of stopped particles by the $\Delta$$E/E$ method.  A
proton is registered when it passes through the inner layer and is
stopped in the second layer, so that the minimum energy of a
proton that can be tracked is determined by the thickness of the
innermost layer.  The maximum energy of tracked protons is given
by the range within the telescope and hence by the total thickness
of all detection layers. Therefore, the primary design goal in the
development of the STT was to combine the thinnest possible
innermost with the thickest available outermost layer of silicon
detector.

A first generation STT~\cite{inti} was already equipped with three
detection layers: a non--structured 60{\ww}$\mu$m thick layer, a
single--sided structured 300{\ww}$\mu$m, and a 5100{\ww}$\mu$m
thick detector. Although serving mainly as a prototype system with
limited size and poor tracking capabilities, it allowed us to
study the reactions $pn \to d \pi^\circ$~\cite{inti}, $pn \to d
\omega$~\cite{barsov}, and it was also used as a
polarimeter~\cite{yaschenko}.

The new STT employs the well--established thickness of the first
generation system, but overcomes the limited tracking capabilities
by using double--sided micro--structured detectors.  The angular
coverage in the forward hemisphere was small, because the position
of the STT with respect to the target was optimized to detect slow
recoil protons in the backward hemisphere. Nevertheless, protons
emitted at angles from about $75^{\circ}$ to $80^{\circ}$ were
unambiguously identified in the STT in coincidence with
elastically scattered deuterons in the FD.

The new STT facilitates $\Delta$$E/E$ proton identification from 2.5
up to 40$\,$MeV with an energy resolution of 150--250$\,$keV
(FWHM). Particle tracking is possible over a wide range of
energies with an angular resolution varying from 1$^\circ$ to
6$^\circ$ (FWHM). The resolution is limited by angular straggling
within the detectors and therefore depends on particle type,
energy, and track inclination. The geometrical limit is defined by
the strip pitch (ranging from 400 to 666{\ww}$\mu$m) and the
distances between the detectors. The STT has self--triggering
capabilities. It identifies a particle passage within 100{\ww}ns
and provides the possibility for fast timing coincidences with
other detector components of the ANKE spectro\-meter, whereby
accidental coincidences can be suppressed significantly. The high
rate capabilities of the STT will be especially important for the
upcoming polarization experiments~\cite{SPIN}, because then two or
more STTs have to be placed in the forward hemisphere. The recent
development of very thick ($>10${\ww}mm) double--sided
micro--structured Si(Li) detectors will allow us to extend further
the accessible energy range of the STT~\cite{protic}.

\section{Measurements with ANKE}\label{Anke_measurements}
\subsection{Identification of Nuclear Reactions \label{chap:identification}}
\begin{figure}[htb]
\begin{center}
\centerline{\epsfxsize=\columnwidth\epsfbox{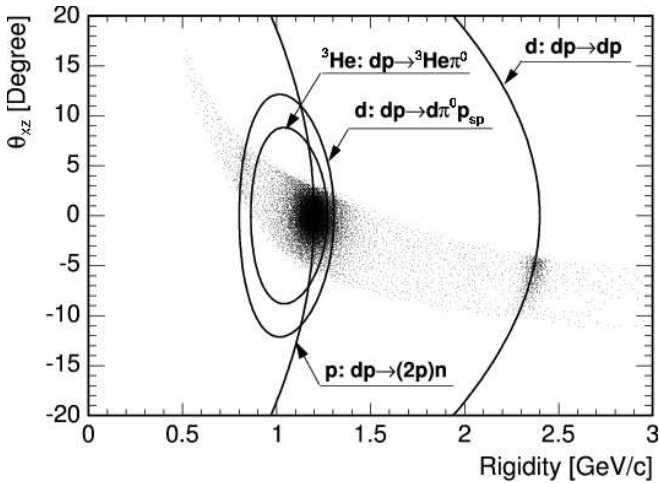}}
\caption{ANKE experimental acceptance for four nuclear reactions of
interest at a deuteron momentum of $p_d=2400${\ww}MeV/c.\label{accep}}
\end{center}
\end{figure}

Figure~\ref{accep} shows the ANKE experimental acceptance for
charged particles as function of the laboratory production
angle and magnetic rigidity. From the loci of the kinematics of
the four reactions that we investigated in this polarization study
it was seen that all of them had reasonable acceptances over some
angular domain.

The main trigger used in the experiment consisted of a coincidence
of different layers in the hodoscope of the FD. The $^3$He were
identified by means of a special energy loss trigger in the FD. In
parallel, self--triggering of the STT was employed to identify
unambiguously $dp$ elastic scattering.

\begin{figure}[htb]
\begin{center}
\centerline{\epsfxsize=\columnwidth\epsfbox{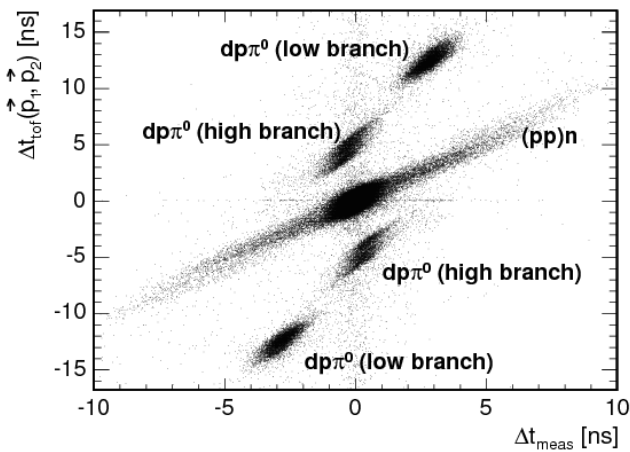}}
\caption{Correlation of the measured time difference $\Delta t_{\rm
meas}$ and the calculated $\Delta
t_{\mathrm{tof}}(\vec{p_1},\vec{p_2})$), assuming that the particle
tracks correspond to protons. \label{tof}}
\end{center}
\end{figure}

Candidate events for different reaction channels can be identified
in a plot of the time of flight difference between target and
hodoscope ($\Delta t_{\rm meas}$) vs the calculated time of flight
difference ($\Delta t_{\mathrm{tof}}(\vec{p_1},\vec{p_2})$),
assuming the two forward particles hitting different hodoscope
counters are protons, as shown in Fig.~\ref{tof}. Real proton
pairs from the charge--exchange breakup $dp \to (pp)n$ are located
along the diagonal of the scatter plot, where for illustration, it
also shows how events from other reactions are transformed by this
procedure.

\subsection{Extraction of the Analyzing Powers}

After classifying the events shown in Figs.~\ref{accep} and
\ref{tof} in terms of various nuclear reactions, these were 
binned in intervals of center--of--mass polar angle
($\theta_{cm}$), or equivalently the three--momentum transfer
($q$), and azimuthal angle ($\phi$). The resulting counts
$N(\theta,\phi)$ or $N(q,\phi)$ were corrected for the
dead--time of the DAQ system, which was typically 10--15\%, and
which was measured with a precision of  better than 1\%.

In order to establish  the relative integrated luminosity of each
of the polarization states involved, we normalized the data using
the beam current information. The signal from the beam current
transformer (BCT) was fed into a voltage--to--frequency converter
within the EDDA electronics and transformed into an optical signal
to avoid deterioration. This information was transported to ANKE
where, after conversion back to a NIM--signal, it was fed into the
ANKE scaler system. In this way the BCT signal was available in
the normal ANKE data stream for each of the polarization states.
The BCT signal is known to about 1\%.

After further correction of the measured counts for luminosity,
with the help of the above beam current information, the various
analyzing powers of the reactions were extracted by fitting
Eq.~(\ref{eq1}) simultaneously to the data obtained for all the
source states, which have different values of the beam
polarizations $P_{z}$ and $P_{zz}$. The details of such a fit in
the $\vec{d}p\to (pp)n$ case are discussed extensively in
Ref.~\cite{Chil1}.

Reliable values of $A_{xx}$ can only be extracted provided that
the apparatus has useful $\phi$--acceptance away from $\phi=0$ or
$180^{\circ}$, but there are significant differences in this for
the four reactions presented below.

\subsection{$\bmath{\vec{d}p\to\,^{3}}$He$\,\bmath{\pi^0}$ Reaction\label{3He}}
\begin{figure}[htb]
\begin{center}
\centerline{\epsfxsize=\columnwidth\epsfbox{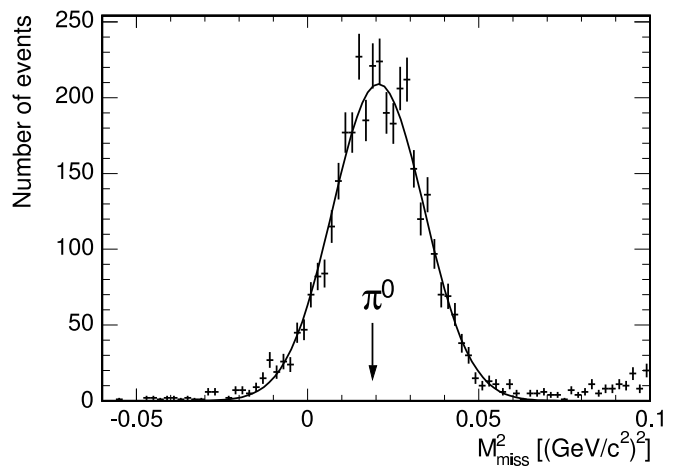}} \caption{
Missing--mass--squared for the $dp\to\,^3\textrm{He}\,X$ reaction
showing a Gaussian fit to a clearly identified $\pi^0$
peak.\label{he3} }
\end{center}
\end{figure}

It is seen from Fig.~\ref{accep} that there is a large acceptance
for the $dp\to\,^3\textrm{He}\,\pi^0$ reaction when the $^3$He are
emitted very close to the initial beam direction. In this region
there are very detailed measurements of the sole non--vanishing
deuteron (tensor) analyzing power $A_{yy}$ as a function of
energy~\cite{Kerboul}. The high--momentum branch of
$^3\textrm{He}$ particles could be selected in off--line analysis
by applying two--dimensional cuts in $\Delta E$ vs momentum and
$\Delta t$ vs momentum for individual layers of the forward
hodoscope. The $\pi^0$ was identified through the missing mass
derived from the $^3\textrm{He}$ measurement, as described in
Ref.~\cite{wronska}. The mean value of the missing mass
distribution (Fig.~\ref{he3}) was close to the pion mass, with
stable background of less than 3\%.  Though the resulting peak has
a large width, this is not critical since, apart from the
radiative capture, there should be no physical background in this
region, and no significant amount is seen in the figure.

Using the counts within a $\pm 2.5 \sigma$ missing--mass
range of the peak in Fig.~\ref{he3}, we
find an analyzing power of $A_{yy}(\theta=0^\circ) = 0.461 \pm
0.030$, where the statistical and systematical uncertainties in
the EDDA beam polarizations given in Table~\ref{tableLEPEDDA} have
not been included. Interpolation of the SATURNE data to our energy
leads to a value of
$A_{yy}(\theta=0^\circ)=0.458\pm0.014$~\cite{Kerboul}.

\subsection{Quasi--Free $\bmath{\vec{n}p\to d\pi^0}$ Scattering \label{dipi0}}
In the bulk of reactions involving collisions with a deuteron at
intermediate energies the process is driven by the interaction
with either the proton or neutron in the nucleus. The other
particle is a \textit{spectator}, a fact that has led to the
extensive use of deuterium as a replacement for a free neutron
target. In the case of a deuteron beam, a \textit{spectator}
proton ($p_{sp}$) would have roughly half the momentum of the beam
and be detectable over a range of angles, as shown in
Fig.~\ref{accep}.
\begin{figure}[htb]
\begin{center}
\centerline{\epsfxsize=\columnwidth\epsfbox{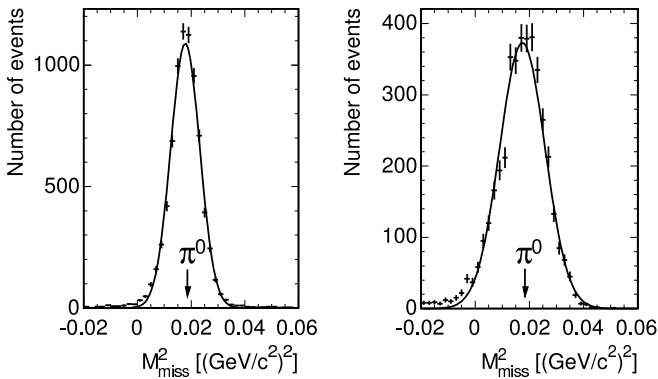}}
\caption{ Missing--mass--squared for the $dp\to p_{sp}d X$
reaction for the slow deuteron (left panel) and fast deuteron
(right panel) branches of the kinematics. In both cases Gaussian
fits to the data are indicated. Events falling within
$\pm2.5\sigma$ of the centers were retained in the analysis.
\label{npdpion} }
\end{center}
\end{figure}


The first step in extracting quasi--free $\vec{n}p\to d\pi^0$
events from our data is to choose two--track events on the basis
of the MWPC information. The momentum vectors were determined with
the help of the magnetic field map of the spectrometer, assuming a
point--like source placed in the center of the beam--target
interaction region. The smallness of the FD solid angle acceptance
leads to a kinematic correlation for events with two or three
particles in the final state. The $\vec{d}p\to p_{sp}d X$
candidates can be clearly identified from the correlation of the
measured time difference $\Delta t_{\rm meas}$ and the calculated
time of flight difference $\Delta
t_{\mathrm{tof}}(\vec{p_1},\vec{p_2})$ (see Fig.~\ref{tof}). The
reaction shows up as isolated regions in a two--dimensional plot
without having to identify  positively the deuterons beforehand.
For both the high deuteron momentum part (forward production in
the cm system), and the low momentum region (backward production)
the missing masses corresponding to the unobserved $\pi^0$ are
clearly seen in Fig.~\ref{npdpion}.

For small deuteron cm angles the spectrometer provides useful
$\phi$--acceptance over the full angular range. However, for
events in the backward hemisphere, this is restricted rather to
$|\phi|<50^{\circ}$, which is quite sufficient to extract the
vector analyzing power $A_y$.

Provided that the spectator momentum in the deuteron rest frame is
small, the vector polarization of the deuteron is completely given
by that of the constituent nucleons. The shape of the spectator
momentum distribution follows the expectation for a reasonable
deuteron wave function~\cite{paris}. By selecting events below
60{\ww}MeV/c, dilution of the neutron polarization due to
$D$--state effects in the deuteron becomes negligible.

\begin{figure}[htb]
\begin{center}
\centerline{\epsfxsize=\columnwidth\epsfbox{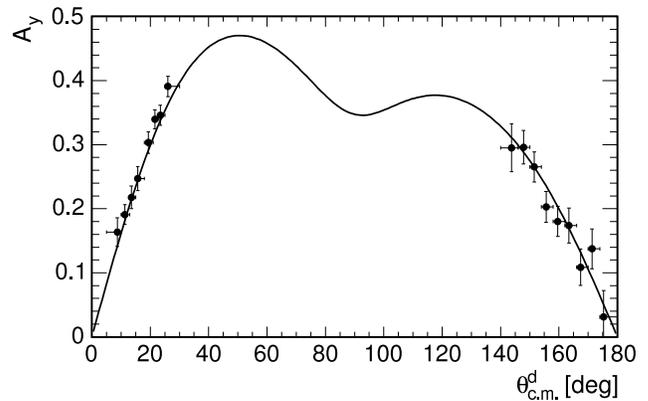}}
\end{center}
\caption{\label{dpi0fig}Analyzing power of the $\vec{n}p\to d\pi^0$
reaction measured at ANKE compared to the curve of values of $A_y$ in
$\vec{p}p\to d\pi^+$, as extracted from the SAID data base~\cite{SAID}
(for numerical values, see Table~\ref{tableallresults}.) }
\end{figure}
Due to isospin invariance, the neutron analyzing power in the
$\vec{n}p\to d\pi^0$ reaction should be identical to that of the
proton in $\vec{p}p\to d\pi^+$, for which extensive data
compilations are available~\cite{SAID}. As shown in
Fig.~\ref{dpi0fig}, the agreement of our result with the SAID data
base is very good for both small and large deuteron cm angles.
(Numerical values can be found in Table~\ref{tableallresults}.)
This result is therefore consistent with the EDDA measurements of
the vector polarization of the deuteron beam. Within small error
bars, typically 2\%, there is no sign of any effect arising from
the tensor polarization of the deuteron beam. This is as expected
for a quasi--free reaction and provides an extra check on the
systematics.

\subsection{Charge--Exchange Reaction $\vec{d}p\to (pp)n$}
The deuteron charge exchange on hydrogen, $\vec{d}p\to (pp)n$ is
defined to be the reaction when the di--proton emerges with low
excitation energy $E_{pp}$. It has been argued that for small
momentum transfers from the deuteron to the di--proton, the
counting rates should depend sensitively upon the tensor
polarization of the beam and that the relatively large analyzing
powers could be estimated reliably in terms of neutron--proton
elastic amplitudes~\cite{Bugg-Wilkin}. These impulse approximation
predictions were successfully tested at 1.6{\ww}GeV using the
SPES-IV spectrometer at SATURNE~\cite{Sams95a}. Similar
measurements at 200{\ww}MeV and 350{\ww}MeV with the EMRIC
device~\cite{Kox} were equally well described by the model, which
was subsequently extended to include final--state interactions in
several of the $pp$ partial waves~\cite{Carbonell}. The reaction
also provides the basis for the design of the POLDER
polarimeter~\cite{POLDER}, which has been used for the
determination of the polarization of the recoil deuteron in
elastic electron--deuteron scattering at JLab~\cite{JLab}. In
addition to a large analyzing power, the signal for the reaction
of two fast protons emerging both with momenta close to half of
the beam, is very distinctive and in our case these fall within
the acceptance of the FD system (Fig.~\ref{accep}).
\begin{figure}[htb]
\begin{center}
\centerline{\epsfxsize=\columnwidth\epsfbox{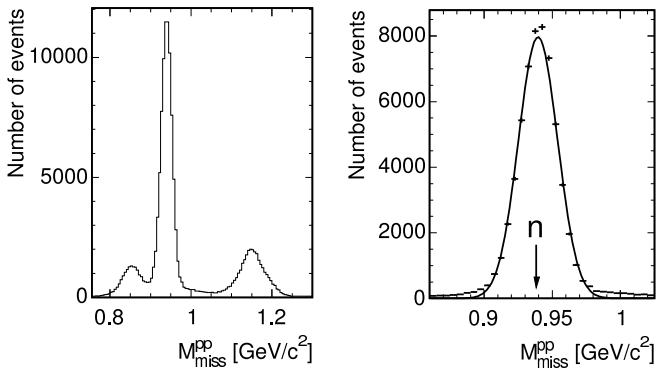}}
\end{center}
\caption{\label{Mpp_distribution} Missing mass distribution of all
observed proton pairs (left). The distribution near the neutron mass
for those pairs selected by the TOF is shown on the right. In this
case there is essentially no background and a Gaussian fit to the
missing mass agrees with that of the neutron to within 1\%.
Events falling within $\pm2.5\sigma$
of the center were retained in the analysis.
}
\end{figure}

\begin{figure}[htb]
\centering
\centerline{\epsfxsize=\columnwidth{\epsfbox{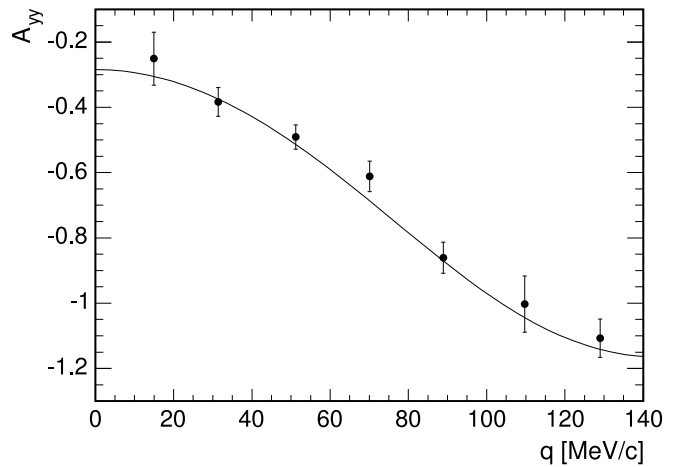}}}
\caption{Values of the tensor analyzing power $A_{yy}$ extracted for
the $\vec{d}p\to (pp)n$ reaction at 1170{\ww}MeV for
$E_{pp}<1${\ww}MeV. The curve corresponds to the predictions of the
impulse approximation of Ref.~\cite{Carbonell}, with input amplitudes
taken from Ref.~\cite{SAID,Igor1}.  } \label{ayy_d2p}
\end{figure}
The detection of proton pairs was already successfully exploited
during earlier measurements~\cite{komarov,yaschenko}.  The
charge--exchange breakup events can be isolated from the scatter
plot in Fig.~\ref{tof} in exactly the same way that the
$p_{sp}d\pi^0$ reaction was studied. The identification of the 
charge--exchange process was finally confirmed from the missing mass with
respect to the observed proton pairs (see
Fig.~\ref{Mpp_distribution}) and the time--difference information.
The spectra for all spin modes reveal a well defined peak at
M$_{\rm miss}^{pp}$ equal to the neutron mass within 1\%. The
background was less than $2\%$ and stable.

As shown in Ref.~\cite{Chil1}, there is sufficiently large acceptance 
for all azimuthal angles when events with $E_{pp}<1${\ww}MeV are
selected. Large values are obtained for both $A_{yy}$ and
$A_{xx}$, whereas $A_y$ is consistent with zero to better than
1\%, as expected from theory~\cite{Bugg-Wilkin}. The $A_{yy}$
results are presented in Fig.~\ref{ayy_d2p} as a function of the
momentum transfer $q$ from the proton to the neutron. They are
compared to the predictions of the impulse approximation, using
the same computer program as that in Ref.~\cite{Carbonell}, though
with modern values of the $np\to pn$ amplitudes taken from the
SAID phase shift analysis~\cite{SAID,Igor1}. Since, for small
$E_{pp}$, this reaction only depends upon the tensor polarization
of the beam~\cite{Bugg-Wilkin}, the agreement shown in Fig.~\ref{ayy_d2p}
is an excellent confirmation of the EDDA values of $P_{zz}$ that
we have used.

\subsection{Deuteron--Proton Elastic Scattering \label{dpelastic}}
\begin{figure}[htb]
\begin{center}
\centerline{
\epsfxsize=\columnwidth\epsfbox{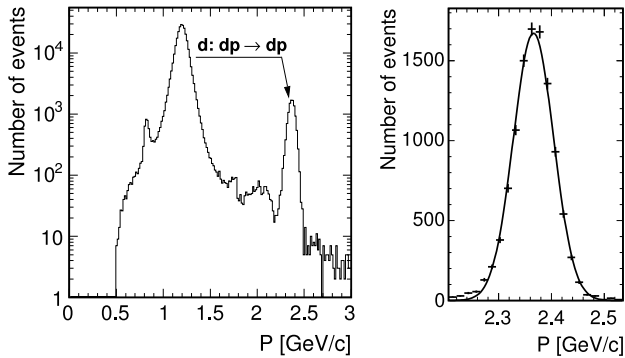}}
\end{center}
\caption{ Left: Single--track momentum spectrum for the $dp$ data
at 2.40{\ww}GeV/c on a logarithmic scale.  Right: Fit result of
the elastic peak region with a Gaussian function on a linear
scale.} \label{fig:p_distribution}
\end{figure}

It is obvious from Fig.~\ref{accep} that deuteron--proton elastic
scattering has a significant acceptance in ANKE for
$4^{\circ}<\theta_\textrm{lab}^d<10^{\circ}$. Because of the very
large cross section, the reaction stands out very clearly in the
momentum and angle--momentum spectra, and is thus easily selected.
The elastic peak region in the momentum spectrum of the single
track events, shown in the left panel of
Fig.~\ref{fig:p_distribution}, was fit by a Gaussian. Values
within $\pm3\sigma$ of the mean momentum were considered to be
good elastic scattering events. An example of such a fit is shown
in the right panel of the figure.

In contrast to the three reactions measured in ANKE that we have
discussed thus far, $\vec{d}p$ elastic scattering depends upon
both the vector and tensor polarizations of the deuteron beam.
Fortunately the analyzing powers $A_{y}$, $A_{yy}$, and $A_{xx}$
of this reaction have been measured at Argonne~\cite{Argonne} for
$T_d=1194${\ww}MeV and SATURNE~\cite{Arvieux} for
$T_d=1198${\ww}MeV.

\begin{figure*}[hbt]
\begin{center}
\subfigure
{\epsfig{file=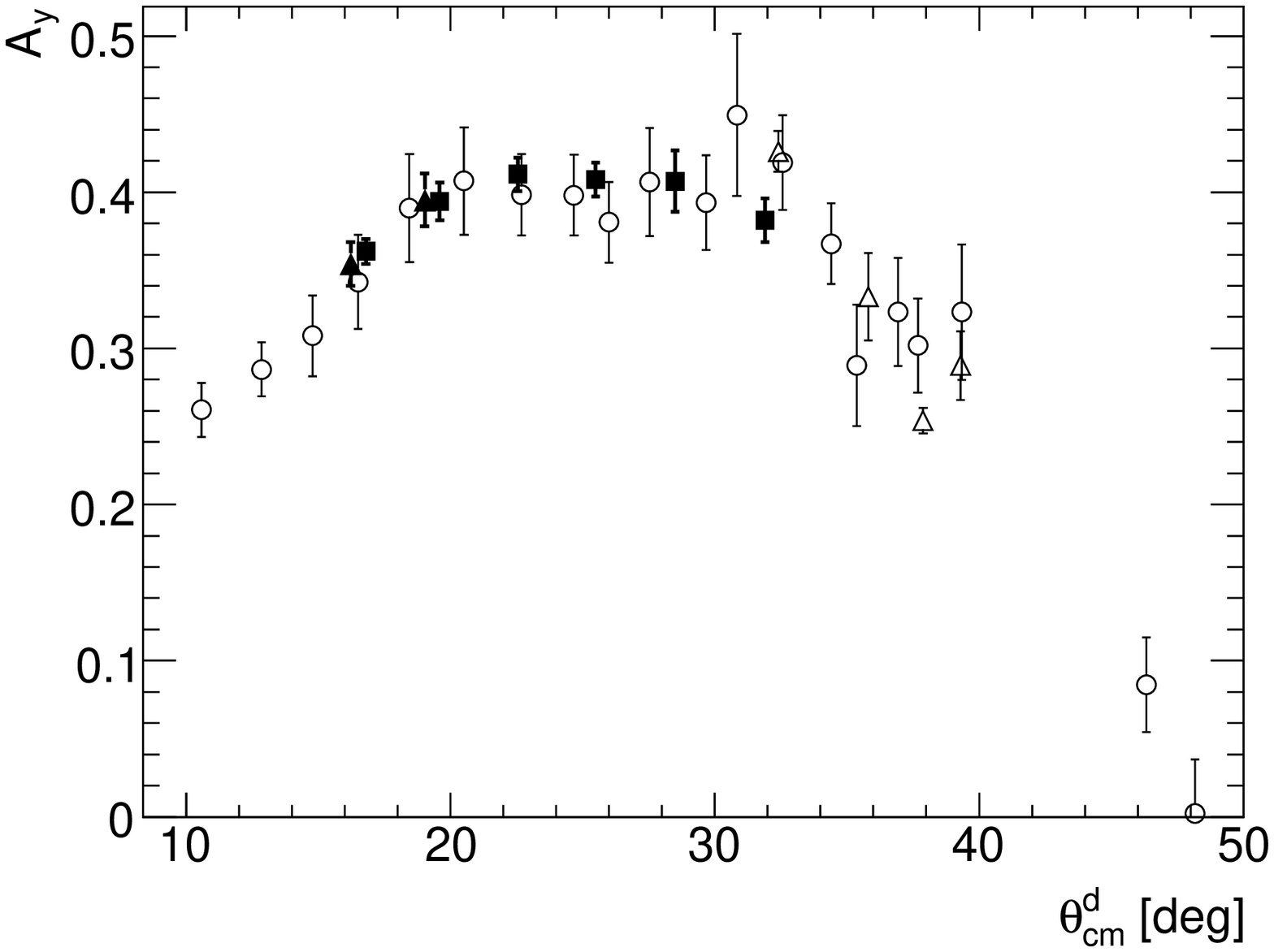,width=\columnwidth}}~~~
\subfigure
{\epsfig{file=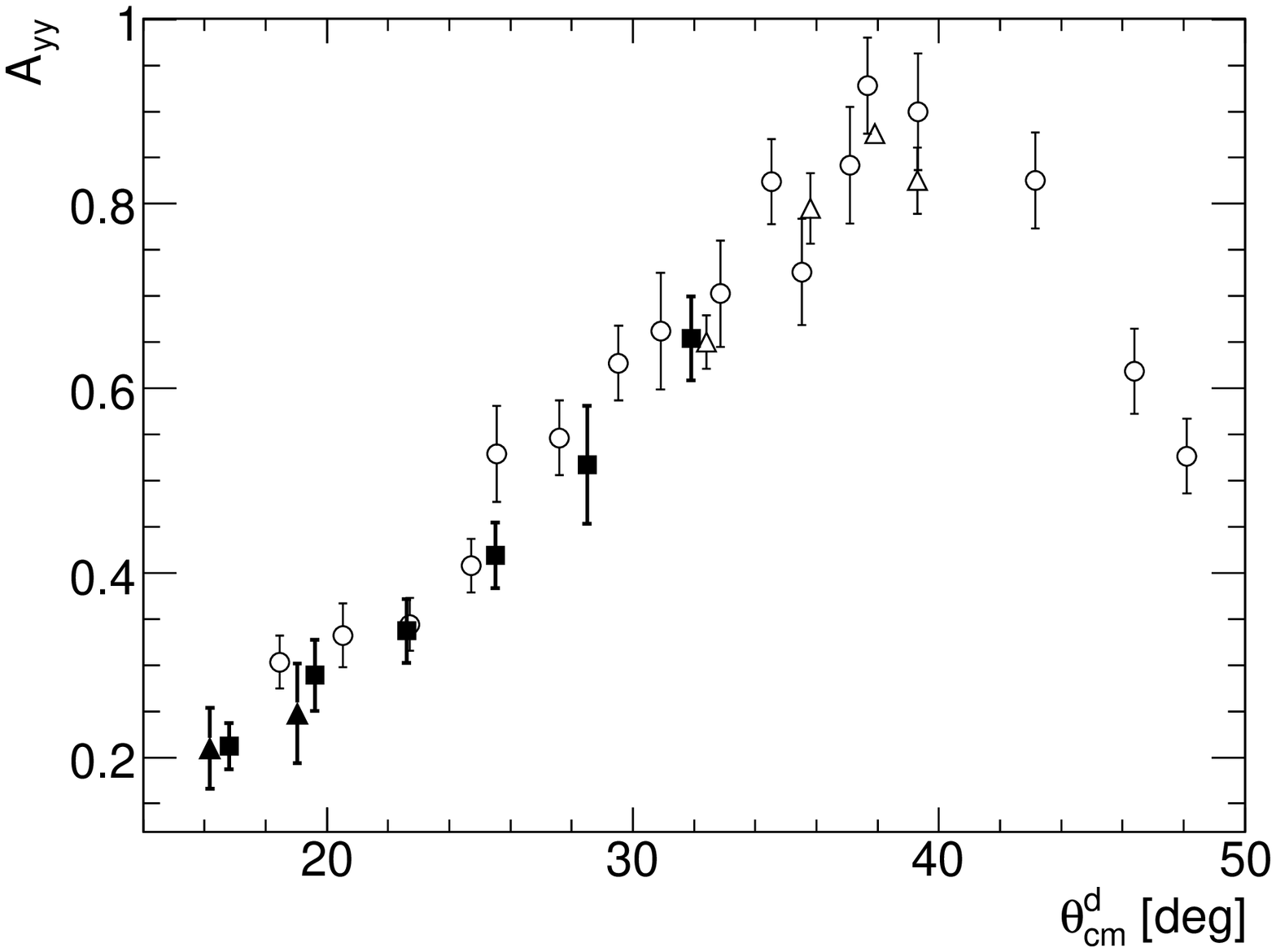,width=\columnwidth}}%
\caption{Vector (left panel) and tensor analyzing powers (right)
for elastic deuteron--proton scattering at small forward angles.
Our data at 1170{\ww}MeV (solid squares) were obtained using
information solely from the forward detector system whereas two
points (solid triangles) resulted from coincidence measurements
with the silicon telescope (for numerical values, see
Table~\ref{tableallresults}). These data are compared to the
results from Argonne at 1194{\ww}MeV~\cite{Argonne} (open circles)
and SATURNE at 1198{\ww}MeV~\cite{Arvieux} (open triangles). It
should be noted that the tensor beam polarization at SATURNE was
the subject of a series of very careful
calibrations~\cite{Arvieux}. \label{Aii_dp}}
\end{center}
\end{figure*}

There is good azimuthal coverage of this reaction in ANKE for
$167^{\circ}<\phi<193^{\circ}$. This is quite sufficient to
extract the values of the vector analyzing power $A_y$ shown in
Fig.~\ref{Aii_dp} (left panel). Comparing with the Argonne and
SATURNE results, the agreement is very good, with all points
coinciding within the published statistical errors.

The situation is not quite as clean in the case of the tensor
analyzing power since the small $\phi$--acceptance does not allow
us to obtain $A_{xx}$. Nevertheless, the finite acceptance does
lead to a small contamination of the $A_{yy}$ measurement from the
$A_{xx}$ term. We therefore introduced a correction of about 4\%
to account for this effect using information derived from the
ratio $A_{xx}/A_{yy}$ determined at Argonne~\cite{Argonne}, where
it should be noted that this ratio does not depend on the beam
polarization used in their analysis. The agreement presented in
Fig.~\ref{Aii_dp} (right panel) is very satisfactory; Numerical
values of $A_y$ and $A_{yy}$ are given in
Table~\ref{tableallresults}.

Though the events identified from the FD information shown in
Fig.~\ref{fig:p_distribution} are very clean, some of the
systematics of the experiment can be checked from the data where
the slow recoil proton from the $\vec{d}p\to dp$ elastic
scattering was detected in the silicon telescope in coincidence
with the deuteron in the FD. Though this restricts both the
acceptance and the statistics, the determination of the angles and
the total lack of any background presents in principle many
advantages. However, as shown in the figure, the results hardly
change when this coincidence is introduced.

\subsection{Precision of the ANKE Results \label{summary}}

The numerical  results from the measurements described in this
section are given in Table~\ref{tableallresults} of Appendix A.
We here discuss separately the precisions with which each of the
reactions determines one of the beam polarizations with the aim of
extracting the best values and errors for $P_z$ and $P_{zz}$ at
$1170\ww{}$MeV. This will also allow us to put limits on the
amount of depolarization by the beam through acceleration to this
energy.

Though the $\vec{d}p\to\,^{3}\textrm{He}\,\pi^0$ reaction in the
forward direction has a very strong tensor analyzing power signal,
the statistical error achieved so far at ANKE does not allow us to
make a strong statement on the basis of these results. Comparing
with the precise results from SATURNE~\cite{Kerboul}, we find that
\begin{equation}
A_{yy}(\textrm{ANKE}) = (1.01\pm0.07)\,A_{yy}(\textrm{SATURNE})\:.
\end{equation}

The $\vec{n}p\to d\pi^0$ reaction is only sensitive to the vector
polarization of the beam. We find that the analyzing power at
different angles is proportional to the SAID prediction for
$\vec{p}p\to d\pi^+$~\cite{SAID} with
\begin{equation}
A_{y}(\textrm{ANKE}) = (1.03\pm0.02)\,A_{y}(\textrm{SAID})\:,
\end{equation}
where $\chi^2/\textit{ndf}=10.5/16$. Although the SAID database
does not allow one to extract errors, the numerous experiments in
this range suggest an overall precision of about
3\%~\cite{Furget}. Allowing also for a possible small violation of
charge independence that links the $\vec{n}p\to d\pi^0$ and
$\vec{p}p\to d\pi^+$ analyzing powers, a very conservative
estimate on the error in $P_z$ from this reaction is about 5\%.

The statistical precision that can be achieved for the tensor
analyzing powers in the $\vec{d}p\to (pp)n$ reaction is very high.
Assuming the validity of the impulse approximation
predictions~\cite{Carbonell} illustrated in Fig.~\ref{ayy_d2p}, we
obtain
\begin{equation}
A_{yy/xx}(\textrm{ANKE}) = (0.98\pm 0.02)\,A_{yy/xx}(\textrm{Theory})\:.
\end{equation}

Since there are some uncertainties in the theoretical model as
well as in the $np$ input, a more cautious limit on the tensor
polarization of the beam would be 5\%. However, it should be noted
that some of the uncertainty might cancel if the reaction were
used to extract information about the $np$ amplitudes at other
energies.

Elastic deuteron--proton scattering is sensitive to the vector and
tensor polarizations of the beam. Comparing our measurements of
$A_y$ for $\vec{d}p\to dp$ with those of Argonne~\cite{Argonne}
shown in Fig.~\ref{Aii_dp}a, the average over the points near the
maximum yields
\begin{equation}
A_y(\textrm{ANKE}) = (1.00\pm 0.03)\,A_y(\textrm{Argonne})\:.
\end{equation}

Unlike the case for the vector analyzing power of $\vec{d}p\to
dp$, there are clear discrepancies between the measurements of
Argonne~\cite{Argonne} and SATURNE~\cite{Arvieux} for the tensor
analyzing power $A_{yy}$ shown in Fig.~\ref{Aii_dp}b, with the
latter being $6\%\pm 3\%$ lower. This was remarked upon in the
SATURNE paper and great care was then taken to establish very
accurate values of the beam polarizations. Using the SATURNE and
renormalized Argonne values, we find for this reaction that
\begin{equation}
A_{yy}(\textrm{ANKE}) = (0.99\pm 0.06)\,A_{yy}(\textrm{SATURNE})\:.
\end{equation}
However, for neither of the two analyzing powers have we tried to
include corrections for the small differences in beam energy
between the different experiments.

Putting all these results together, we see that
\begin{eqnarray}
\nonumber
A_{y}(\textrm{ANKE})&=&(1.01\pm 0.03)\,A_{y}(\textrm{Expected})\:,\\
A_{yy}(\textrm{ANKE})&=&(0.99\pm 0.03)\,A_{yy}(\textrm{Expected})\:.
\label{final}
\end{eqnarray}
The error bars on the ``Expected'' results are obtained from
theory and a variety of experiments around $1170\ww{}$MeV.
However, they do not explicitly include the uncertainties of
$2.1\%$ and $2.6\%$ in the SATURNE values of $P_z$ and $P_{zz}$.
If one takes these into account then the uncertainties in the
vector and tensor polarizations of the deuteron beam in ANKE are
both on the 4\% level.

The central values shown in Eq.~(\ref{final}) reflect the possible
loss of polarization during the acceleration of the deuterons from
the EDDA energy to that of ANKE. Though these indicate very little
depolarization, one cannot draw very tight limits on this effect
because of the uncertainties introduced by the calibration of the
EDDA polarimeter. Taking just the systematic errors of 4\% here,
we suggest that any polarization loss is below 6\% for both the
vector and tensor parameters.

\section{Conclusion and Outlook\label{chap:conclusions}}
By measuring five analyzing powers in ANKE, we have shown that it
is possible to determine both the vector and tensor polarizations
of the deuteron beam at 1170{\ww}MeV with precisions of about 4\%
each. The vector and tensor polarizations are typically about 74\%
and 59\%, respectively, of the ideal values that could be provided
by the source. Taking these results in conjunction with the values
of the polarizations measured with the EDDA polarimeter, we find
no evidence for any depolarization in the acceleration from EDDA
to ANKE. However, we can only put limits of about 6\% on such
effects because of the extra uncertainty in the absolute
calibration of EDDA. A somewhat stronger limit might be found if
we assumed that any degradation of $P_{zz}$ were associated with
one of $P_{z}$, though this argument should be taken with some
caution~\cite{morozov}. One way to eliminate calibration effects
completely would be by decelerating the beam and remeasuring in
EDDA itself.

Since ANKE can accept many nuclear reactions simultaneously there
are, of course, other processes that might be used to check the
beam polarizations. One of these is proton--deuteron backward
elastic scattering, for which tensor analyzing powers are
available~\cite{A-B}. However, the cross section is rather low and
our measurement of the analyzing power $A_{yy}$ has a 15\%
statistical error, which make the reaction of very limited use for
the present purposes. One might think of using quasi--elastic
proton--proton scattering to determine $P_z$, with one particle
being detected in ANKE and the other in the STT. Though we have
high statistics measurements of this reaction, it is not
appropriate to use this to calibrate the vector polarization. In
the small momentum transfer region allowed by the ANKE acceptance,
there are large corrections to the quasi--free picture coming from
final--state interactions~\cite{Chil1}, so that one cannot use
free $pp$ scattering to determine the polarization. Nevertheless
the reaction might be useful to provide on--line monitoring of
$P_z$.

The results achieved here allow us to extend our $\vec{d}p\to
(pp)n$ measurements to higher energy, where they can be used to
add to the existing $np$ scattering data base. It is known that in
the forward direction $A_{yy}$ provides similar information to
that of the spin--transfer parameter $K_{0nn0}$ in neutron--proton
elastic scattering in the backward direction~\cite{Bugg-Wilkin}.
At 788\ww{}MeV this parameter has been measured with a
systematic precision of better than 3\% but the statistical error
bars were at the 8--15\% level~\cite{McNaughton}. Nevertheless,
these data represent a significant contribution to the $np$
amplitude analysis at this energy. Hence, the measurements of
$A_{yy}$ and $A_{xx}$ in the $\vec{d}p\to (pp)n$ reaction with
systematic errors below 5\% should provide useful data at this and
especially at higher energies, where the existing information is
poorer.

The deuteron charge exchange is just one part of the polarization
schedule at the COSY--ANKE facility~\cite{SPIN}. Polarized
hydrogen and deuterium targets are currently being commissioned,
and the combination of polarized beams and targets will lead
to a very rich physics program.\\

\begin{acknowledgments}
We would like to thank the accelerator crew for the operation of COSY,
the injector cyclotron, and the polarized ion source.  We are grateful
to I.I.~Strakovsky for providing us with up--to--date neutron--proton
amplitudes, and to E.J.~Stephenson for valuable information regarding
deuteron analyzing powers at low energies. We thank D.~Proti\'{c} and
Th.~Krings (FZJ) for providing the Si(Li) detectors.
\end{acknowledgments}
\begin{appendix}
\section{Tabulated Results}
In Table~\ref{tableallresults}, we give the numerical results of
the measurements of analyzing powers $A_y$ and $A_{yy}$ of
$\vec{d}p$ elastic scattering (Sec.~\ref{dpelastic}), analyzing
power $A_y$ of the quasi--free $\vec{n}p\to d\pi^0$ reaction
(Sec.~\ref{dipi0}), and the value of the analyzing power $A_{yy}$
of the $\vec{d}p \to ^3$He$\,\pi^0$ reaction from ANKE
(Sec.~\ref{3He}), together with the previously obtained SATURNE
result~\cite{Kerboul}.
\begin{table}[htb]
\renewcommand{\arraystretch}{1.2}
\begin{small}
\begin{center}
\begin{tabular}{|c|c|c|c|}
\hline
$\vec{d}p \to dp       $   & $\theta_{cm}^{d} $ & $A_{y}$ & $A_{yy}$   \\
\hline
FD & $\phantom{0}16.8^{\circ}$&$0.362\pm0.008$&$0.212\pm0.025$  \\
                  & $\phantom{0}19.6^{\circ}$&$0.394\pm0.012$&$0.289\pm0.039$  \\
                  & $\phantom{0}22.6^{\circ}$&$0.410\pm0.011$&$0.337\pm0.034$  \\
                  & $\phantom{0}25.5^{\circ}$&$0.408\pm0.011$&$0.419\pm0.035$  \\
                  & $\phantom{0}28.5^{\circ}$&$0.407\pm0.020$&$0.517\pm0.064$  \\
                  & $\phantom{0}31.9^{\circ}$&$0.382\pm0.014$&$0.654\pm0.045$  \\
\hline
FD+STT          &$\phantom{0}16.2^{\circ}$&$0.354\pm0.014$&$0.209\pm0.044$  \\
                      &$\phantom{0}19.1^{\circ}$&$0.395\pm0.017$&$0.248\pm0.054$  \\\hline\hline
  $\vec{n}p \to d\pi^{0}$  & $\theta_{cm}^{d}$ & $A_y$ &$A_y(\textrm{SAID})$ \\
\hline
 & $ \phantom{0}\phantom{0}8.7^{\circ}$  & $0.164\pm0.023$  &$0.142$\\
                        & $\phantom{0}11.3^{\circ}$  & $0.191\pm0.015$ &$0.181$\\
                        & $\phantom{0}13.5^{\circ}$  & $0.218\pm0.018$ &$0.214$\\
                        & $\phantom{0}15.7^{\circ}$  & $0.247\pm0.019$ &$0.244$\\
                        & $\phantom{0}19.4^{\circ}$  & $0.303\pm0.017$ &$0.292$\\
                        & $\phantom{0}21.6^{\circ}$  & $0.340\pm0.015$ &$0.318$\\
                        & $\phantom{0}23.4^{\circ}$  & $0.346\pm0.016$ &$0.337$\\
                        & $\phantom{0}26.1^{\circ}$  & $0.391\pm0.016$ &$0.364$\\\hline
                        & $143.8^{\circ}$ & $0.295\pm0.037$ &$0.311$ \\
                        & $147.8^{\circ}$ & $0.296\pm0.026$ &$0.289$ \\
                        & $151.6^{\circ}$ & $0.265\pm0.024$ &$0.265$ \\
                        & $155.6^{\circ}$ & $0.203\pm0.024$ &$0.236$ \\
                        & $159.6^{\circ}$ & $0.180\pm0.023$ &$0.203$ \\
                        & $163.4^{\circ}$ & $0.174\pm0.027$ &$0.170$ \\
                        & $167.4^{\circ}$ & $0.109\pm0.028$ &$0.132$ \\
                        & $171.3^{\circ}$ & $0.137\pm0.031$ &$0.092$ \\
                        & $175.3^{\circ}$ & $0.031\pm0.041$ &$0.050$ \\ \hline\hline
  $\vec{d}p \to ^3$He$\,\pi^0$ & $\theta_{cm}^{^3{\rm He}}$ & &  $A_{yy}$    \\\hline
 ANKE    & $0^\circ$        & & $0.461 \pm 0.030$\\
 SATURNE & $0^\circ$        & & $0.458 \pm 0.014$\\\hline
\end{tabular}
\end{center}
\end{small}
\caption{Analyzing powers $A_y$ and $A_{yy}$ of the $\vec{d}p\to
dp$ reaction as function of $\theta_{cm}^{d}$ (top, using only the
FD, or a coincidence of FD and STT), analyzing power $A_y$ of the
$\vec{n}p\to d\pi^0$ reaction as function of $\theta_{cm}^{d}$
(middle), and the values of the analyzing power $A_{yy}$ of the
$\vec{d}p \to ^3$He$\,\pi^0$ reaction at $\theta_{cm}^{^3{\rm
He}}=0^\circ$ from ANKE and SATURNE~\cite{Kerboul} (bottom). Also
shown are the SAID predictions for $A_y(\vec{p}p\to d\pi^+)$
obtained using the SP96 solution~\cite{SAID}.
\label{tableallresults}}
\end{table}
\end{appendix}
\clearpage



\begin{thebibliography}{99}
\bibitem{cosyring} A floorplan of the COSY accelerator complex can be
found in Fig.~1 of ref.~\cite{morozov}.
\bibitem{rf-switch} The time it takes for the atomic beam polarization
to stabilize after a change of state is very short compared to the
cycle time. When only the rf is switched in an RFT, the atomic beam
polarization becomes stable within 0.02{\ww}s, and within 0.5{\ww}s,
if in addition the magnetic field is changed.
\bibitem{improvement}In order to improve the polarimetry directly at
the COSY ion source, a new polarimeter for proton and deuteron beams,
based on the concept of the ANKE Lamb--shift
polarimeter~\cite{engels}, is presently being installed.
\end{thebibliography}
\end{document}